\documentclass[11pt]{article} 
\usepackage{blindtext}
\usepackage[a4paper, total={6in, 8in}]{geometry}

\usepackage{url}

\usepackage{amsmath,amssymb,amstext}
\usepackage{xcolor}
\usepackage[flushleft]{threeparttable}

\usepackage{extdash}

\definecolor{linkColor}{RGB}{0,70,120}

\usepackage[colorlinks=true,allcolors=linkColor,pdfborder={0 0 0},pdfencoding=auto]{hyperref}

\usepackage{graphicx}
\usepackage[format=plain,%
            labelsep=period,%
            font={small,sf},%
            width=0.85\textwidth,%
            labelfont=bf,%
            figurename=Fig.%
            ]{caption}

\usepackage[detect-all]{siunitx}

\definecolor{myGreen}{RGB}{0,100,0}
\definecolor{myTeal}{RGB}{0,80,150}
\definecolor{darkred}{RGB}{200,0,0}

\usepackage[normalem]{ulem}

\renewcommand{\eqref}[1]{\textup{{\normalfont Eq.~(\ref{#1}}\normalfont)}}

\begin{document}
\begin{flushleft}

{\Large \bfseries Bridging scales in a multiscale pattern-forming system}

\bigskip
{
Laeschkir~W\"urthner\textsuperscript{1,*}, 
Fridtjof~Brauns\textsuperscript{1,2,*}, 
Grzegorz~Pawlik\textsuperscript{3},
Jacob~Halatek\textsuperscript{1,4},
Jacob~Kerssemakers\textsuperscript{3},
Cees~Dekker\textsuperscript{3,\textdagger},
and Erwin~Frey\textsuperscript{1,5,\textdagger}
}

\bigskip
\small\textsuperscript{1}Arnold Sommerfeld Center for Theoretical Physics and Center for NanoScience, Department of Physics, Ludwig-Maximilians-Universit\"at M\"unchen, Theresienstra\ss e 37, D-80333 M\"unchen, Germany

\medskip
\small\textsuperscript{2}Present address: Kavli Institute for Theoretical Physics, University of California Santa Barbara, Santa Barbara, CA 93106, USA

\medskip
\small\textsuperscript{3}Department of Bionanoscience, Kavli Institute of Nanoscience Delft, Delft University of Technology, Van der Maasweg 9, 2629 HZ Delft, the Netherlands

\medskip
\small\textsuperscript{4}Research Department, Oxford BioMedica Ltd., Windrush Court, Transport Way, Oxford, OX4 6LT, UK

\medskip
\small\textsuperscript{5}Max Planck School Matter to Life, Hofgartenstraße 8, D-80539 Munich, Germany

\bigskip
\bigskip

\small\textsuperscript{*}L.W. and F.B.\ contributed equally to this work. \\
\small\textsuperscript{\textdagger}Corresponding authors. Email: frey@lmu.de or c.dekker@tudelft.nl

\end{flushleft}
\bigskip 

\begin{abstract}
\small Self-organized pattern formation is vital for many biological processes. 
Reaction-diffusion models have advanced our understanding of how biological systems develop spatial structures, starting from homogeneity. 
However, biological processes inherently involve multiple spatial and temporal scales and transition from one pattern to another over time, rather than progressing from homogeneity to a pattern.
To deal with such multiscale systems, coarse-graining methods are needed that allow the dynamics to be reduced to the relevant degrees of freedom at large scales, but without losing information about the patterns at the small scales.
Here, we present a semi-phenomenological approach which exploits mass-conservation in pattern formation, and enables to reconstruct information about patterns from the large-scale dynamics.
The basic idea is to partition the domain into distinct regions (coarse-grain) and determine instantaneous dispersion relations in each region, which ultimately inform about local pattern-forming instabilities.
We illustrate our approach by studying the Min system, a paradigmatic model for protein pattern formation.
By performing simulations, we first show that the Min system produces multiscale patterns in a spatially heterogeneous geometry.
This prediction is confirmed experimentally by in vitro reconstitution of the Min system.
Using a recently developed theoretical framework for mass-conserving reaction-diffusion systems, we show that the spatiotemporal evolution of the total protein densities on large-scales reliably predicts the pattern-forming dynamics.
Our approach provides an alternative and versatile theoretical framework for complex systems where analytical coarse-graining methods are not applicable, and can in principle be applied to a wide range of systems with an underlying conservation law. 
\end{abstract}

\clearpage\newpage

\section*{Introduction}
Pattern formation is fundamental for the spatiotemporal organization of biological processes, such as cell division, chemotaxis, and morphogenesis. More than half a century ago, Turing showed theoretically how local interactions (chemical reactions) and diffusion of chemical species can lead to spontaneous spatial patterns \cite{Turing1952}.
Such reaction--diffusion systems have been successfully used to explain pattern formation phenomena in nature that arise self-organized from a stable homogeneous steady state \cite{Kondo.Asai1995, Klausmeier1999, Green.Sharpe2015, Halatek.etal2018}.
The analysis proposed by Turing allows to predict the emergence of patterns with a characteristic length scale as long as the entire dynamics remains in the vicinity of the homogeneous steady state \cite{Cross.Hohenberg1993}.
The validity of Turing's approach has been also tested experimentally for coupled chemical oscillators, and was found to reliably predict the experimental observations, provided that the model parameters are spatially and temporally uniform~\cite{Tompkins.etal2014}.
Pattern-forming systems, however, are generally heterogeneous and therefore far from homogeneity, and involve multiple spatial and temporal scales.
An intriguing example of biological pattern formation is morphogenesis, in which the spatiotemporal patterns of morphogens dictate the future shape of an organism that is orders of magnitude larger than its constituents \cite{Green.Sharpe2015}.
On a smaller scale, protein concentration patterns in cells are essential for the spatiotemporal control of cellular processes such as cell division and motility \cite{Frey.Brauns2020, Halatek.etal2018, Goryachev.Leda2017}.
Protein patterns can exhibit fascinating multiscale characteristics \cite{Glock.etal2019a} and form in hierarchies of patterns on several scales that affect one another \cite{Wigbers.etal2021}.

Such complex multiscale biological processes involve many degrees of freedom at multiple scales, rendering it difficult to analyze them and gain insight into the underlying principles. 
To make progress on this issue, one needs to use systematic coarse-graining schemes that allow the dynamics to be reduced to the essential degrees of freedom at the relevant time and length scales.
For instance, a well-known and powerful method is the renormalization group theory \cite{Taeuber.2014}.
Unfortunately, this method is restricted to the vicinity of critical points.
The Mori-Zwanzig formalism \cite{Zwanzig2001} is another important approach which allows to decompose the dynamics of a system into `fast' and `slow' variables by means of projection operators. One arrives at a closed set of equations for the slow variables, while the fast variables are treated as noise.
One property that these methods have in common is that the scales that have been integrated out or eliminated are not \emph{resolved}, and cannot be recovered from the coarse-grained level of description.
This is most apparent in the Mori-Zwanzig formalism, where the eliminated degrees of freedom appear effectively as noise terms on the resolved scales.
For pattern-forming systems, one is however interested in the patterns on the \textit{unresolved scales}\footnote{We adapt the term unresolved scales from the computational fluid dynamics literature to refer to the (small) scales that have been integrated out in the coarse-grained description.}
as they usually have a specific function in biological systems.
This raises the question of whether it is possible to reconstruct information about the unresolved scales from the dynamics at the resolved scales?
Indeed, amplitude equations describe the long-wavelength amplitude modulations of an underlying short-wavelength base pattern and therefore resolve both the small and the large scales. 
Unfortunately, however, they are limited to the vicinity of the supercritical onset of pattern formation~\cite{Cross.Hohenberg1993} (including weakly subcritical cases) and only feasible in simple geometries where the orthonormal basis functions of the diffusion operator can be found in closed analytical form.
Hence, to fill these gaps, one relies on new concepts to deal with multiscale systems.

Here, we propose a semi-phenomenological approach to overcome these mathematical limitations in the concrete context of mass--conserving reaction--diffusion (MCRD) systems. 
Recently, a new theoretical framework for MCRD systems has been introduced \cite{Halatek.Frey2018, Brauns.etal2020} that allows one to characterize their dynamics in the highly nonlinear regime.
The basic idea is to consider reaction--diffusion system as decomposed into a set of reactive compartments which are spatially coupled by diffusion.
For an isolated compartment, one can determine the steady state (\emph{local equilibrium}) and its stability properties which both depend on the total densities within that compartment.
Since diffusion causes the lateral redistribution of these total densities, these local equilibria will change over time.
This concept of \emph{moving local equilibria} enables one to study the physical mechanisms underlying pattern formation and characterize the dynamics far away from the homogeneous steady state.
The fact that one is able to characterize the dynamics far from homogeneity suggests that the \emph{local equilibria theory} may be a promising approach to study heterogeneous systems. 
We therefore asked whether the ideas from local equilibria theory would be applicable to investigate multiscale patterns?

\begin{figure}[b!]
\centering
\includegraphics[width=.55\textwidth]{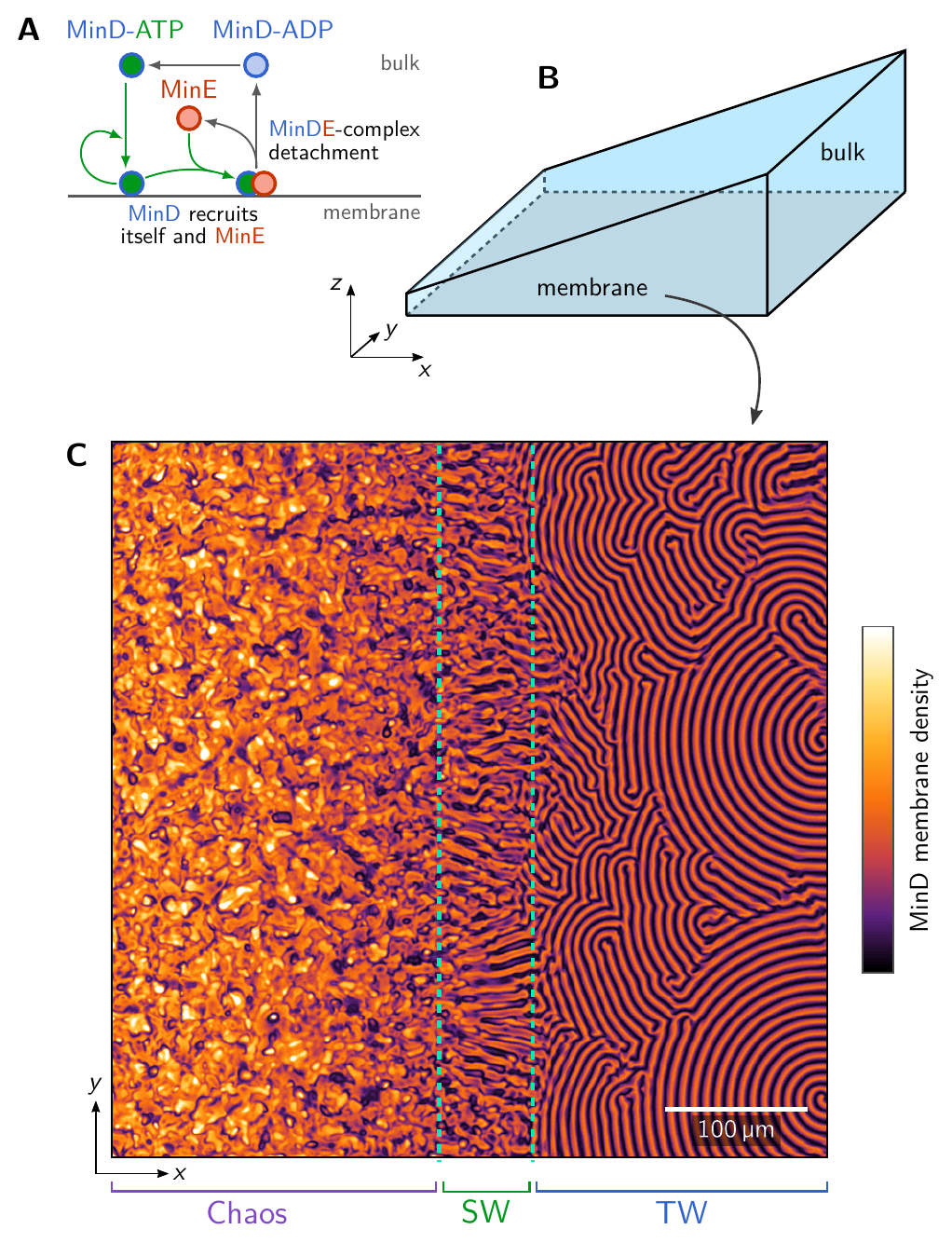}
\caption{(\textit{A}) Schematic illustration of the Min-protein reaction network. 
  (\textit{B}) Wedge-geometry with a membrane surface at the bottom plane ($z=0$) and bulk height $H(x)$ increasing linearly along the $x$ direction.
  (\textit{C}) Snapshot of the membrane-density of MinD, obtained by numerically simulating the Min dynamics Eqs.~\ref{eq:cyt_dyn}--\ref{eq:bc} in the geometry shown in (\textit{B}). One observes regions with chaotic patterns, standing waves (SW, dashed green outline) and traveling waves (TW) along the membrane and at different bulk heights; see Movie~S1.}
\label{fig:setup}
\end{figure}

To pursue this question, we use the Min protein system of \textit{E.\ coli} which has emerged as a paradigmatic model system for the study of pattern formation in cell biology \cite{Adler.etal1967,deBoer.etal1989,Loose.etal2008,Frey.etal2018,Ramm.etal2019}. 
Its dynamics is driven by two proteins, MinD and MinE, which cycle between cytosolic and membrane--bound states and interact nonlinearly on the membrane (Fig.~\ref{fig:setup}A).
In \textit{E.~coli}, these proteins oscillate from cell pole to cell pole and thereby position the cell division machinery to midcell \cite{Adler.etal1967,deBoer.etal1989}.
Studying the Min dynamics in various reconstituted systems has led to the discovery of a rich set of patterns including traveling waves and spirals \cite{Loose.etal2008}, chaotic patterns \cite{Ivanov.Mizuuchi2010,Vecchiarelli.etal2016,Denk.etal2018,Glock.etal2019a}, ``homogeneous pulsing'' \cite{Litschel.etal2018, Godino.etal2019, Kohyama.etal2019}, as well as quasi-stationary labyrinths, spots, and mesh-like patterns \cite{Glock.etal2019a,Glock.etal2019}.
Theoretical analysis of mathematical models has lead to the key insight --- and experimentally confirmed prediction --- that the average total densities of MinD and MinE and the bulk height are key control parameters for pattern formation in the reconstituted Min system \cite{Halatek.etal2018,Brauns.etal2021a}.
The rich set of patterns, experimental accessibility \textit{in vitro} and theoretical understanding make the Min-system an ideal candidate to investigate the role of spatial heterogeneity on pattern formation.

Since varying the bulk height affects the local equilibrium state and is a key control parameter for pattern formation \cite{Halatek.etal2018,  Brauns.etal2021a}, we study the Min dynamics in a wedge--shaped geometry with a membrane placed on the bottom surface (Fig.~\ref{fig:setup}B). 
While there are many distinct ways to introduce large--scale spatial heterogeneities into the system, e.g.\ by introducing space-dependent kinetic rates, 
we chose to use a wedge geometry because it is relatively easy to implement experimentally. 
In numerical simulations, we find that the system exhibits a striking range of transient patterns, that coexist in different spatial regions along the membrane
(Movie~S1 and Fig.~\ref{fig:setup}C). As time progresses, patterns in different regions change and transition to other patterns.

To characterize these complex dynamics that play out on multiple spatial and temporal scales, we generalize the concept of dispersion relations (obtained from a linear stability analysis) by applying it to sections of the domain, which we term \emph{regional dispersion relations}.
Combining this approach with the local equilibria theory \cite{Halatek.Frey2018, Brauns.etal2020, Frey.Brauns2020}, we show that one can reconstruct the type and characteristics of patterns on small scales from the local protein mass densities, which we identify as the essential degrees of freedom on large spatial and temporal scales, i.e.\ the ``hydrodynamic variables'' of the system.
The key to this reconstruction are correlations between the regional pattern characteristics and instantaneous, regional dispersion relations, calculated from the instantaneous regional mass densities.
Over time, these masses change due to diffusive redistribution, resulting in qualitatively different regional dispersion relations that indicate the local pattern type in the system.
This reconstruction of small-scale features (on \textit{unresolved scales}), together with a coarse-grained description for the mass-redistribution dynamics on large scales allows us to understand and predict the long--term temporal evolution of the system.
A major advantage of our approach is that it is based on a linear theory and therefore conceptually and technically simple to apply.

A key prediction from our numerical simulations and theoretical analysis is that different pattern types form at different positions along the wedge shaped geometry.
To test this prediction experimentally, we performed experiments with a reconstituted Min system in wedge-shaped microfluidic cells. 
In agreement with the theoretical prediction, we find a range of transient patterns coexisting in different spatial regions along the membrane.

\section*{Results}
\subsection*{The Min protein system in wedge geometry}
Mathematically, the Min-protein dynamics is described by bulk-surface coupled reaction--diffusion equations, which describe the concentrations of cytosolic proteins MinD-ATP, MinD-ADP, and MinE, $\mathbf{c} = (c_\mathrm{DD},c_\mathrm{DT},c_\mathrm{E})$, in the bulk volume $\mathcal{V}$, and the concentrations of membrane-bound MinD and MinDE complexes, $\mathbf{m} = (m_\mathrm{d},m_\mathrm{de})$, on the surface $\mathcal{S}$.
For the wedge geometry, in spatial coordinates $\mathbf{x} = (x,y,z)$, we place the membrane surface (with lateral dimensions $L\times L$) in the $x{-}y$ plane at $z = 0$ and let the bulk height vary as a linear ramp from $H_0$ to $H_1$ along the $x$-direction (see Fig.~\ref{fig:setup}B).

The dynamics of bulk components $\textbf{c}(\textbf{x},t)$ is governed by the equation
\begin{equation}
    \partial_t \mathbf{c}(\mathbf{x},t)=D_c \nabla^2 \mathbf{c} + \Lambda \mathbf{c},
    \label{eq:cyt_dyn}
\end{equation}
where $D_c$ denotes the bulk diffusion constant and the matrix ${\Lambda = \text{diag}(-\lambda,\lambda,0)}$ describes nucleotide exchange of MinD in the bulk.
The dynamics of membrane components $\mathbf{m}(x,y,t)$ is constrained to the membrane surface and takes the form:
\begin{equation}
   \partial_t \mathbf{m}(x,y,t)=D_m \nabla^2_{\mathcal{S}} \mathbf{m} + \mathbf{r}(\mathbf{c}|_{z = 0},\mathbf{m}),
    \label{eq:mem_dyn}
\end{equation}
where $D_m$ is the membrane diffusion constant and ${\nabla^2_{\mathcal{S}} = \partial_x^2 + \partial_y^2}$ is the surface Laplacian. 
The membrane reactions $\mathbf{r}$, which comprise attachment, detachment, and recruitment processes of Min proteins, are specified in the Materials and Methods section.

The dynamics in the bulk and on the surface are coupled by reactive boundary conditions,
\begin{equation}
   {-}D_c \partial_z \mathbf{c}|_{z=0}=\boldsymbol{f}(\mathbf{c}|_{z=0},\mathbf{m}),
    \label{eq:bc}
\end{equation}
that describe the bulk fluxes induced by attachment and detachment of proteins at the membrane (see Materials and Methods).
At the remaining boundaries, no-flux boundary conditions are imposed such that the system is closed.
Together, the above dynamics conserve the average mass densities of MinD and MinE:
\begin{subequations}
\label{eq:total_avg_densities}
\begin{align}
    \bar{n}_\mathrm{D} \, |\mathcal{V}|
    &=
   \langle m_\mathrm{d} + m_\mathrm{de} \rangle_\mathcal{S}^{} \, |\mathcal{S}| 
    +
    \langle c_\mathrm{D} \,  \rangle_\mathcal{V}^{} |\mathcal{V}| 
    \, , 
    \label{eq:avg_mass_minD}\\
    \bar{n}_\mathrm{E} \, |\mathcal{V}|
    &=
   \langle m_\mathrm{de} \rangle_\mathcal{S}^{} \,  |\mathcal{S}| 
    +
    \langle c_\mathrm{E}\rangle_\mathcal{V}^{} \, |\mathcal{V}| \, ,
\label{eq:avg_mass_minE}
\end{align}
\end{subequations}
where $c_\mathrm{D} = c_\mathrm{DD} + c_\mathrm{DT}$ is the total cytosolic MinD concentration; $\langle \cdot \rangle_\mathcal{S}^{}$ and $\langle \cdot \rangle_\mathcal{V}^{}$ denote the mean on the surface and in the bulk respectively; $|\mathcal{S}|$ and $|\mathcal{V}|$ are the total surface area and bulk volume (see Materials and Methods).

Using finite element (FEM) simulations we investigated the spatiotemporal dynamics of the Min system in wedge geometry.
Our simulations show a broad range of different patterns --- including traveling waves, standing waves and chaotic patterns --- coexisting in different spatial regions of the membrane (see Movie~S1 and Fig.~\ref{fig:setup}C). 
Interestingly, the regions where these patterns are found change over time as the patterns transition from one type to another.
For long simulation times, we observe that patterns transition to standing waves, such that the entire domain is covered by a single pattern type in the final steady state. 
The pattern in steady state depends on the specific choice of parameters, and therefore can be altered by changing the model parameters (Fig.~S1 and Movie~S2).

\subsection*{Experimental implementation}
\begin{figure*}[b!]
\centering
\includegraphics[width=1\linewidth]{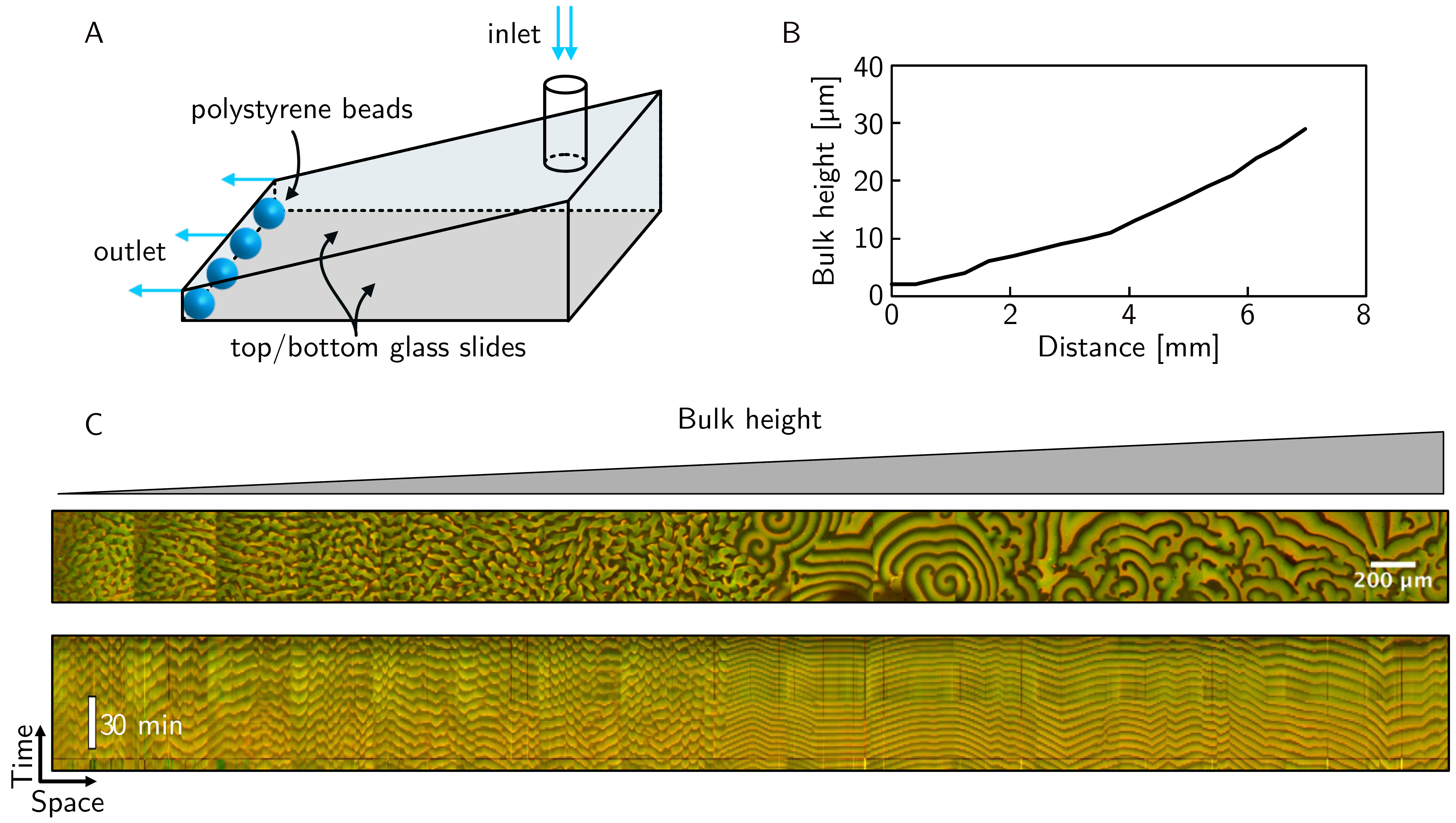}
\caption{Experimentally observed Min patterns in a wedge flow cell. (\textit{A}) Schematic presentation of the experimental setup. Both, the bottom and the top surface (glass slides) are covered with a lipid bilayer. (\textit{B}) Measurement of the bulk height profile of the flow cell versus distance along the lateral length of the wedge. The height was measured microscopically by z-stacks at multiple spots. (\textit{C}) Snapshot of the Min pattern along the wedge, the picture was obtained by stitching individual adjacent images. Shown is a merge of MinD (green) and MinE (red) channels. The bottom figure shows a kymograph for intensities taken along the center line in the top figure.
}
\label{fig:exp-setup}
\end{figure*}

We tested our theoretical prediction on this multi-scale dynamics in an experimental system consisting of a wedge-shaped microfluidic flow chamber (Fig.~\ref{fig:exp-setup}A).
The bottom and top surface of the wedge were covered with a supported lipid bilayer consisting of DOPG:DOPC (30:70 \%) which mimics the natural membrane composition of E. coli~\cite{Vecchiarelli.etal2014}.
The length of the wedge was typically about $\SIrange[range-phrase=-,range-units=single]{8}{14}{\milli m}$ and the width about $\SIrange[range-phrase=-,range-units=single]{3}{4}{\milli m}$. 
The bulk height range was approximately $\SIrange[range-phrase=-,range-units=single]{2}{50}{\micro m}$ (Fig.~\ref{fig:exp-setup}B).
Min proteins were distributed in the chamber by rapid injection of a solution containing $\SI{1}{\micro M}$ MinD and $\SI{1}{\micro M}$ MinE (including 10 \% fluorescently labelled MinD and MinE proteins for visualization), together with $\SI{5}{\milli M}$ ATP and an ATP-regeneration system~\cite{Brauns.etal2021a}.

Figure~\ref{fig:exp-setup}C shows a snapshot of Min protein patterns along the bottom surface of the wedge geometry 30 minutes after injection. 
The experiments exhibit the same essential hallmarks of multiscale Min protein patterns that we observed in our numerical simulations. 
In particular, consistent with our simulations, we observe a sequence of distinct spatiotemporal patterns coexisting in different spatial regions of the membrane (Fig.~\ref{fig:exp-setup}C and Movie~S3): 
At regions of low bulk height (approximately between $\SIrange[range-phrase=-,range-units=single]{2}{10}{\micro m}$), one typically observes chaotic patterns and standing waves, whereas traveling wave patterns emerge at regions of large bulk height ($>\SI{10}{\micro m}$). 
Furthermore, as in the simulation, we observe a sharp boundary between regions that contain traveling wave patterns and regions that contain rather chaotic and standing wave patterns, and this boundary establishes quickly within a few minutes (Fig.~S2 and Movie~S4).
Overall, the observations provide a striking verification of the height-dependent patterns predicted in the simulations. 

There are also some differences between the patterns in the experiment and in our numerical simulations. 
First, while we observed occasional transitions from one pattern into another in our experiments (Fig.~S3 and Movie~S5), these transitions occurred frequently and were more pronounced in the simulations. 
This is explained by the lateral length of the experimental setup, that is about an order of magnitude larger as compared to the simulation setup, which is the main reason why we observe more frequent transitions between different patterns in the simulations, as will become clear later. 
Second, in contrast to the simulations, we noticed some homogeneous oscillations in the experiments, which are characterized by large (homogeneous) density patches on the membrane (typically few hundred micrometers in size) that oscillate with time (Figs.~S3--~S4 and Movies~S5--~S7). 
We attribute this difference to the following: Due to the fabrication method of the microfluidic flow chamber, both the bottom and top surface of the wedge were covered with a supported lipid bilayer. 
In recent work, it has been shown that membrane-to-membrane crosstalk (i.e., between top and bottom surface) is responsible for the emergence of homogeneous oscillations~\cite{Brauns.etal2021a}.
In our simulations, however, we assume that Min proteins can only bind to the bottom membrane, which explains why we do not observe homogeneous oscillations. 

Taken together, we have a system that exhibits a fascinatingly rich transient dynamics and involves patterns and transitions between them on multiple spatial and temporal scales. We are therefore left with the key question: Can we explain the cause why different patterns form in different spatial regions and how they transition from one to another over time? Moreover, is it possible to identify and reduce the system to its essential degrees of freedom?
A standard way to address these questions mathematically would be to perform a multiscale analysis and to derive amplitude equations that describe the large-scale spatiotemporal evolution of the pattern amplitudes~\cite{Cross.Hohenberg1993}.
This would greatly simplify the problem as it allows to obtain a quantitative relationship between the small-scale patterns and the large-scale dynamics (slowly varying pattern amplitudes), thus ultimately enabling one to reconstruct the patterns from the reduced dynamics at large length and time scales~\cite{Matthews.Cox2000, Cox.Matthews2003, Winterbottom.etal2005, Winterbottom.etal2008}.
Carrying out this analysis requires determining the set of orthogonal eigenmodes for the diffusion operator that satisfy the boundary conditions. 
In a one-dimensional domain, these eigenmodes are simply Fourier modes. 
Unfortunately, in the wedge geometry with bulk-surface coupling, the eigenmodes can not be found analytically, thus precluding the use of the amplitude equation framework.
Moreover, amplitude equations are restricted to the vicinity of supercritical and weakly subcritical bifurcations \cite{Cross.Hohenberg1993,Aranson.Kramer2002}.
The Min patterns we observe here, however, are generically subcritical~\cite{Brauns.etal2020} and exhibit large amplitudes~\cite{Halatek.Frey2018,Brauns.etal2021a}.
We therefore aim to develop a new approach that overcomes these restrictions.

\subsection*{Instantaneous, regional dispersion relations predict patterns}
 
The analysis of pattern-forming systems usually starts with calculating the homogeneous steady state (HSS) solutions and performing a linear stability analysis around these states. 
This yields a \emph{dispersion relation} that informs about the growth rate $\sigma(q)$ of small spatial perturbations with a certain wavenumber $q$.
However, the dispersion relation is generally only informative in the vicinity of the homogeneous steady state~\cite{Turing1952,Cross.Hohenberg1993}, and thus unreliable for large amplitude patterns. 
Moreover, the spatial variation of parameters even precludes the existence of a global HSS, so that a global dispersion relation can no longer be determined.
To overcome these limitations, we adopt a semi-phenomenological approach where we generalize the concept of dispersion relations.

\begin{figure*}[b!]
\centering
\includegraphics[width=.9\linewidth]{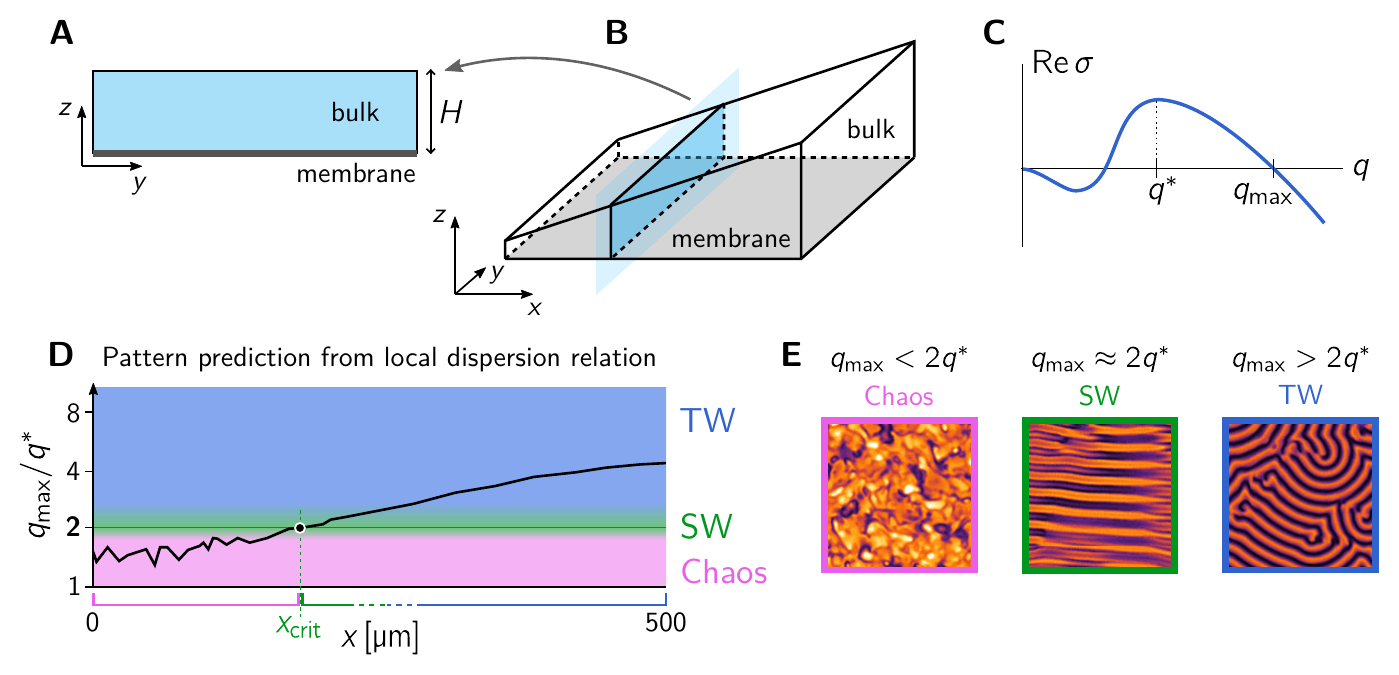}
\caption{(\textit{A}) Rectangular geometry with membrane at the bottom edge representing a slice through the three-dimensional in vitro system.
  (\textit{B}) A slice through the wedge geometry. For each such slice, at a given instance in time, we calculate the instantaneous total densities, averaged along its length $\langle \tilde{n}_\mathrm{D,E} \rangle_y (t,x)$, from the numerical simulation data. From these slice-averaged total densities, we can then calculate the corresponding local homogeneous steady state and its dispersion relation. 
  (\textit{C}) \emph{Dispersion relation} with fastest growing mode $q^*$ and right edge of the band of unstable modes $q_\text{max}$ indicated. The ratio $q_\text{max}/q^*$ has been empirically found to correlate with the type of the fully developed pattern, with a sharp transition from chaotic patterns for $q_\text{max}/q^* < 2$ to ordered patterns for $q_\text{max}/q^* > 2$. Close to the transition, standing waves are found, while travelling waves form for larger ratios $q_\text{max}/q^*$ \cite{Halatek.Frey2018}.
  (\textit{D}) Mode ratio $q_\text{max}/q^*$ as a function of the slice position $x$ for a given instance in time. The background shading indicates the type of pattern expected from the ``commensurability criterion.''
  (\textit{E}) Representative snapshots of the three distinct pattern types: spatiotemporal chaos, standing waves (SW) and traveling waves (TW).
  }
\label{fig:slice-analysis}
\end{figure*}

Let us consider the wedge as dissected into a collection of two-dimensional slices along the direction of constant bulk height.
Each slice corresponds to a rectangular geometry with a bulk height that depends on the position of the slice in the wedge (see Fig.~\ref{fig:slice-analysis}A,B).
Next, for each slice and at each point in time, we calculate instantaneous total densities of MinD and MinE, averaged over the slice length $\langle \tilde{n}_\mathrm{D,E} \rangle_y (t,x)$ (Materials and Methods).
The average total densities, together with the local bulk height $H(x)$, then serve as parameters for the \emph{regional dispersion relation} in each slice
\begin{equation}
    \sigma\left(q; H(x), \langle \tilde{n}_\mathrm{D,E} \rangle_y (t,x)\right),
    \label{eq:regional_dispersion_relation}
\end{equation}
which is straightforward to determine because the slice represents a rectangular geometry~\cite{Halatek.Frey2018, Denk.etal2018, Brauns.etal2021a} (see Fig.~\ref{fig:slice-analysis}A,B and Supplementary Information).
While the bulk height $H(x)$ varies linearly in space, the average total densities $\langle \tilde{n}_\mathrm{D,E} \rangle_y (t,x)$
are dynamic quantities and depend on the slice position $x$ as well as on time $t$, since the diffusive coupling between the slices redistributes mass. 
It follows that the regional dispersion relation depends on the spatial position and is dynamic: $\sigma(q; x,t)$.
This generalizes classical dispersion relations, which are by definition independent of space and time.

How does this spatially and temporally varying dispersion relation inform about the system's dynamics?
As in uniform systems that exhibit homogeneous steady states, it serves as a criterion for the onset of pattern formation and for estimating the characteristic wavelength of the initial pattern that is formed.
While these insights are generally limited to the linear regime~\cite{Turing1952,Cross.Hohenberg1993}, recent theoretical findings for the Min system in a two-dimensional rectangular geometry (representing a slice geometry) have shown that the dispersion relation reliably predicts the pattern type in the fully nonlinear regime~\cite{Halatek.etal2018}.
In particular, it was shown that depending on the total densities of Min proteins, $\bar{n}_{\text{D}}$ and $\bar{n}_{\text{E}}$, and the bulk height $H$, the system exhibits a variety of different patterns on the membrane, such as chaos, standing waves, and traveling waves 
\cite{Halatek.Frey2018,Brauns.etal2021a}.
Moreover, a careful analysis of numerical simulations has interestingly revealed a strong one-to-one correlation between the dispersion relation and the fully developed patterns in the highly nonlinear regime~\cite{Halatek.Frey2018}:
A \textit{commensurability criterion} between the unstable mode with the shortest wavelength $q_{\text{max}}$ and the fastest growing mode $q^*$ has been found that determines the pattern type (Fig.~\ref{fig:slice-analysis}C--E).
In short, it has been shown that $q_{\text{max}}/q^* < 2$ coincides with the regime of chemical turbulence (spatiotemporal chaos), whereas for $q_{\text{max}}/q^* > 2$ the system exhibits ordered patterns (standing/traveling waves).
Standing wave patterns are found close to the commensurability transition $q_{\text{max}}/q^* \gtrsim 2$, while traveling waves are found further away from the threshold.
In the following, we use this observed one-to-one correspondence between the dispersion relation and the fully developed patterns to reconstruct the small scale pattern types from coarse grained densities.

\begin{figure*}[b!]
\centering
\includegraphics[width=.8\linewidth]{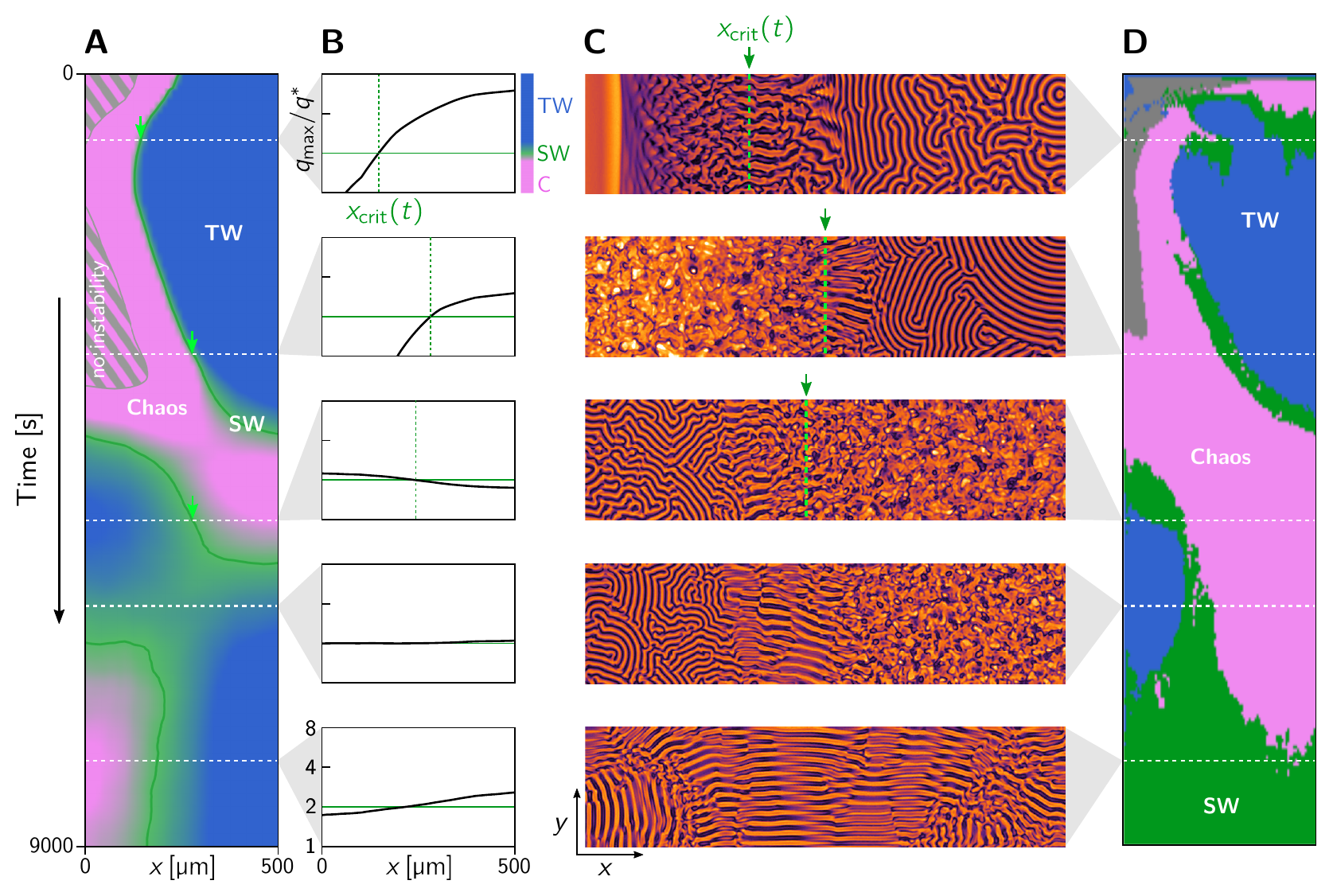}
\caption{(\textit{A}) Kymograph showing the pattern-type prediction from the commensurability criterion (cf.\ Fig.~\ref{fig:slice-analysis}D). The green line shows $x_\mathrm{crit}(t)$ where $q_\text{max}/q^* = 2$, indicating the transition from chaotic to ordered patterns. Green arrows mark the position $x_\mathrm{crit}(t)$ for the times indicated by dashed white lines.
  (\textit{B}) Plots of the mode ratio $q_\text{max}/q^*$, determined from the local dispersion relation, as a function of spatial position $x$ for several representative times (dashed white lines in (A)). In the second to last row, the entire domain is near the critical ratio $q_\text{max}/q^* = 2$, predicting the global emergence of standing waves (see last row).
  (\textit{C}) Snapshots of the membrane patterns (MinD density, cf.\ Fig.~\ref{fig:setup}) from the full numerical simulation. The green dashed line indicates $x_\mathrm{crit}(t)$. Note the standing wave patterns found near $x_\mathrm{crit}(t)$. Their fronts are aligned along the bulk height gradient such that the sequence of wavenodes lies on lines of constant bulk height.
  (\textit{D}) Machine-learning based pattern classification using \textit{ilastik}~\cite{Berg.etal2019} (see Materials and Methods).
  }
\label{fig:pattern-type-kymograph}
\end{figure*}

To that end, we extracted the average total densities in each slice as a function of time from the numerical simulation.
Based on these densities we then calculated the instantaneous regional dispersion relation in each slice and extracted the ratio $q_{\text{max}}/q^*$ as a function of slice position $x$ and time $t$ (Fig.~\ref{fig:slice-analysis}C--E).
The resulting pattern-type prediction is shown in the space-time plot (kymograph) in Fig.~\ref{fig:pattern-type-kymograph}A.
Figure~\ref{fig:pattern-type-kymograph}B shows the ratio $q_{\text{max}}/q^*$ as a function of slice position $x$ for a set of representative times (cf.\ Fig.~\ref{fig:slice-analysis}D). 
The pattern-type prediction Fig.~\ref{fig:pattern-type-kymograph}A is then obtained from these ratios via the mapping shown in  Fig.~\ref{fig:slice-analysis}D,E.

We find that this prediction correlates well with the patterns observed in the full numerical simulation (Fig.~\ref{fig:pattern-type-kymograph}C,D and Movie~S8).
In particular, the temporally changing position $x_\mathrm{crit}(t)$, marking regions where $q_{\text{max}}/q^* = 2$ (indicated by the green arrows and dashed lines in Fig.~\ref{fig:pattern-type-kymograph}B and C), agrees with the position along the wedge where traveling wave patterns transition to chaotic patterns. 
In the vicinity of $x_\mathrm{crit}(t)$ we observe a band of standing waves as expected from the ``commensurability criterion'' \cite{Halatek.Frey2018}.
While the transition from chaos to order at $q_{\text{max}}/q^* = 2$ is sharp, we were not able to identify a sharp criterion for the transition from standing waves to traveling waves. 
Since the ratio $q_{\text{max}}/q^*$ and with it $x_\mathrm{crit}(t)$ are entirely determined by the slice-averaged masses $\langle \tilde{n}_\mathrm{D,E} \rangle_y (x,t)$, we conclude that these masses are the essential degrees of freedom of the system at large scales.

Notably, we find that there are slight differences between the predictions and the actual patterns for large times (see Fig.~\ref{fig:pattern-type-kymograph}A--C).
The reason for these deviations lies in the model parameters, which were chosen such that the entire domain is near the critical mode ratio $q_{\text{max}}/q^* = 2$ for large times. 
This renders the dynamics, and the prediction from the regional dispersion relation highly sensitive to slight variations of the regional total masses. 
Hence, the fact that our method is still able to qualitatively predict the dynamics in this case underscores the robustness of our approach.
In the Supplemental Information, we provide additional results where the parameters were chosen such that the mode ratio is deep in the traveling wave regime ($q_{\text{max}}/q^* > 2)$ for late times. 
In this case, we obtain an excellent agreement between our predictions and the patterns observed in the numerical simulations (see Fig.~S1).

Next, we ask whether one can find an approximate coarse-grained dynamics for these redistributed masses.
Such a description would enable us to predict the time evolution of the redistributed masses independently from the full numerical simulations. 
One can then use the commensurability criterion to predict the pattern types that will form in different spatial regions as a function of the redistributed masses.
In the next section we will show how one can find such a description.

\subsection*{Large-scale dynamics is driven by redistribution of mass}
In general, mass redistribution between different spatial regions of the wedge is caused by diffusive fluxes due to concentration gradients.
Similar as in the previous section, we consider here the redistribution of mass between slices along the wedge (Fig.~\ref{fig:slice-analysis}B).
Since membrane diffusion is by two orders of magnitude slower than bulk diffusion it may be neglected, such that redistribution of protein mass between slices is governed by bulk diffusion alone (Materials and Methods)
\begin{equation} 
\label{eq:mass-redistri-exact-driftterm}
    \partial_t \langle n_i \rangle_{y,z} (x,t)
    \approx
    D_c
   \langle \partial_x^2 c_i \rangle_{y,z}
    +
    D_c 
    \tfrac{\partial_x H(x)}{H(x)}
    \langle \partial_x  c_i \rangle_{y,z},
\end{equation}
$\text{for } i = \mathrm{D,E}$.
Here, the second term accounts for the spatial variation of the bulk height, and thus the different volumes of neighboring slices between which the diffusive flux $D_c \langle \partial_x c_i \rangle_{y,z}$ redistributes mass. This can be seen by rewriting \eqref{eq:mass-redistri-exact-driftterm} in the form of a continuity equation
\begin{equation} \label{eq:mass-redistri-exact}
     \partial_t \big[ H(x) \cdot \langle n_i \rangle_{y,z} (x,t) \big]
    \approx 
    - \partial_x \big[
    H(x) \cdot J^\mathrm{diff}_i
    \big]
\end{equation}
with the diffusive fluxes given by $J^\mathrm{diff}_i := - D_c \langle \partial_x c_i \rangle_{y,z}$.
Since the area of slices increases along the positive $x-\text{direction}$, the diffusive fluxes $J^\mathrm{diff}_i$ on the right-hand side of \eqref{eq:mass-redistri-exact} are rescaled by the bulk height $H(x)$. 
These equations seem to be simple, but unfortunately they are not closed, since the slice-averaged cytosolic densities $\langle c_i \rangle_{y,z}(x,t)$ appear on the right hand side.

\begin{figure}[b!]
\centering
\includegraphics[width=0.65\linewidth]{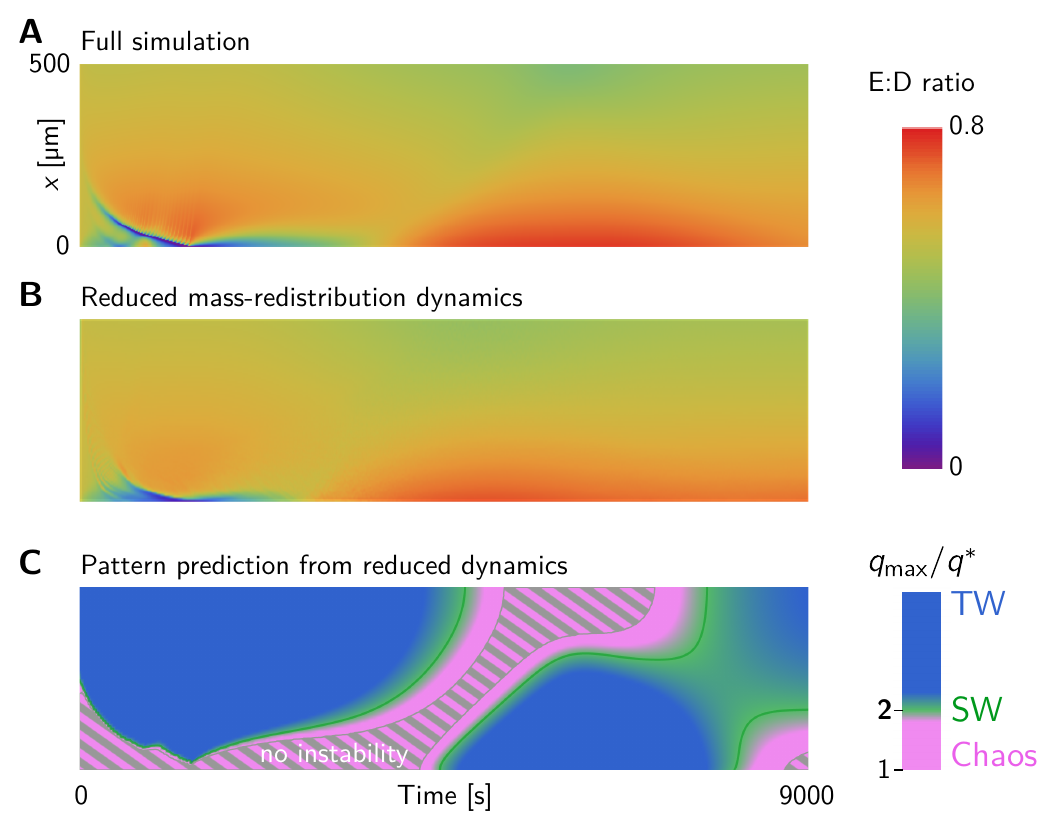}
\caption{(\textit{A},\textit{B}) Kymographs showing the total-density ratio of MinE to MinD (E:D ratio) from the full numerical simulation (\textit{A}) and from local-equilibria based reduced dynamics (\textit{B}).
  (\textit{C}) Kymograph showing the pattern-type prediction using the commensurability criterion based on the total densities from the reduced dynamics. Note the excellent qualitative agreement to the pattern-type prediction based on total densities from the full numerical simulation in Fig.~\ref{fig:pattern-type-kymograph}A.}
\label{fig:proxy-sim}
\end{figure}

We are interested in the dynamics of $\langle n_i \rangle_{y,z}$ on timescales much longer than typical oscillation periods of the patterns.
Therefore, following the intuition gained from previous works on MCRD systems \cite{Brauns.etal2020,Brauns.etal2021b}, we assume that one can approximate the slice-averaged cytosol concentrations by the homogeneous steady-state concentration in each slice
\begin{equation} \label{eq:LQSSA}
    \langle c_i \rangle_{y,z}(x,t)
    \approx c_i^*(x,t) :=
    c_i^*\bigl(
        H(x),
        \langle n_\mathrm{D} \rangle_{y,z},
        \langle n_\mathrm{E} \rangle_{y,z}
    \bigr).
\end{equation}
This assumes that the spatial average over many wavelengths in $y$-direction is well approximated by the instantaneous homogeneous steady state in a slice.
These steady state concentrations only depend on the slices bulk height $H(x)$ and the slice-averaged total densities $\langle n_i\rangle_{y,z} (x,t)$.
Thus, the above approximation yields a closed set of equations for the mass-densities
\begin{equation} \label{eq:mass-redistri-approx}
    \partial_t \langle n_i \rangle_{y,z} (x,t) \approx
    D_c \partial_x^2 c^*_i(x,t)
    +
    D_c 
    \tfrac{\partial_x H(x)}{H(x)} 
    \partial_x  c^*_i(x,t).
\end{equation}
We will call this the \emph{reduced dynamics} in the following.
Since the homogeneous steady states may also undergo a saddle-node bifurcation, characterized by the emergence of three steady states (two stable, one unstable), this may lead to discontinuities in $c^*_i$. To regularize the dynamics, $c_i$ is not set identical to $c_i^*$ but relaxes towards it on a fast timescale (see SI for details).

Given the initial densities $\langle n_i \rangle_{y,z}(x,0)$, one can numerically solve the reduced dynamics \eqref{eq:mass-redistri-approx} to predict the entire time evolution of the slice-averaged masses and hence the dispersion relation at each point along the $x-\text{direction}$. 
Figure~\ref{fig:proxy-sim}C shows the regional pattern types predicted from the reduced dynamics.
We find good qualitative agreement for the distribution and transition of patterns as observed in the numerical simulations (cf.\ Fig.~\ref{fig:pattern-type-kymograph}A). 
The main difference to the full numerical simulations is a slight quantitative deviation in the timescale, where the dynamics predicted by~\eqref{eq:mass-redistri-approx} is slightly slower compared to the full numerical simulation.
We also note that the reduced dynamics predicts a larger region of no instabilities as compared to the numerical simulations (cf. Figs.~\ref{fig:pattern-type-kymograph}A and~\ref{fig:proxy-sim}C).
This is because the chaotic regime is rather narrow and close to the regime for which the dispersion relation predicts no instability (cf.\ Figs.~\ref{fig:slice-analysis}D and~\ref{fig:pattern-type-kymograph}B).
In addition, since the patterns emerge from a subcritical bifurcation \cite{Halatek.Frey2018} (a generic property of mass-conserving systems~\cite{Brauns.etal2020}), large amplitude patterns can be excited and maintained even below the instability threshold.

Figure~\ref{fig:proxy-sim}A,B compare the time evolution of the slice-averaged total densities from the full numerical simulation and the solution obtained from the reduced dynamics. 
The colors in the kymographs indicate the total density ratio of MinE and MinD (short, E:D ratio), which is a key control parameter in the Min-protein dynamics \cite{Halatek.Frey2018}.

\section*{Discussion}

Multiscale patterns in biological systems often emerge from hierarchical systems,
which are organized in a modular fashion. 
Each level of the hierarchy instructs dynamics on the next level which operates on a smaller spatial scale. 
For instance, along developmental trajectories of many organisms, upstream patterns such as maternal gradients instruct downstream gene-expression patterns on increasingly smaller scales \cite{Petkova.etal2019,Wigbers.etal2021}. 
Importantly, on each level of the hierarchy, there is a clean separation between (spatially varying) control parameters and dynamical variables.

In contrast, in the system we have studied here, there is no such separation as the globally conserved total densities play a dual role: they are both dynamical variables and act as control parameters \cite{Halatek.Frey2018,Brauns.etal2020}.
Building on this key feature has allowed us to explain and predict the intriguingly complex patterns found in large-scale numerical simulations.
The values of the total densities of MinD and MinE locally control the pattern type: we showed that a ``regional dispersion relation'' calculated from the regional average densities reliably predicts the pattern type. 
At the same time, concentration gradients in the bulk drive mass redistribution of MinD and MinE. 
Therefore, the total densities are hydrodynamic variables on large scales which control pattern formation on small scales. 
This separation of scales enabled us to derive a reduced dynamics for the total densities on large spatial and temporal scales which predicts the long-term dynamics of the system.

Notably, the dual role of total densities as dynamic variables and control parameters also plays out at the small scale of the patterns themselves \cite{Halatek.Frey2018,Brauns.etal2020}. 
Here, instantaneous \emph{local} total densities control \emph{local} equilibria and their stability, which serve as proxies for the local dynamics. The local dynamics cause gradients, which drive diffusive redistribution of the total densities---in turn causing changes in the local dynamics.
In the Min system, this point of view has led to a detailed understanding of the emergence of chaos near onset and of the transition to standing and traveling waves \cite{Halatek.Frey2018}.  
From a general perspective, the concept of local equilibria controlled by total local densities is at the core of a number of recent theoretical advances in the field of mass-conserving, pattern-forming systems \cite{Brauns.etal2020,Brauns.etal2021b,Brauns.etal2021c,Frey.Brauns2020}.

In addition to the dynamically changing total densities, the bulk height is also a (fixed) heterogeneous control parameter in our system.
The bulk height (or more generally volume-to-surface ratio) is an important control parameter for bulk-surface coupled pattern-forming systems~\cite{Halatek.Frey2018, Brauns.etal2021a}. 
Here, the bulk height gradient of the wedge serves to induce spatiotemporal heterogeneities in the total densities.
Alternatively, one could induce heterogeneities in the total densities via spatial gradients of the kinetic rates or by imposing a heterogeneous initial condition in the total densities.
However, these alternatives are difficult to realize experimentally in a reproducible and controlled manner, which is the main reason why we chose the wedge setup in this work.
In a third scenario, large-scale gradients in the densities may also emerge spontaneously and be maintained in the absence of ``external'' heterogeneities. 

\begin{figure}[h!]
    \centering
    \includegraphics[width=0.7\linewidth]{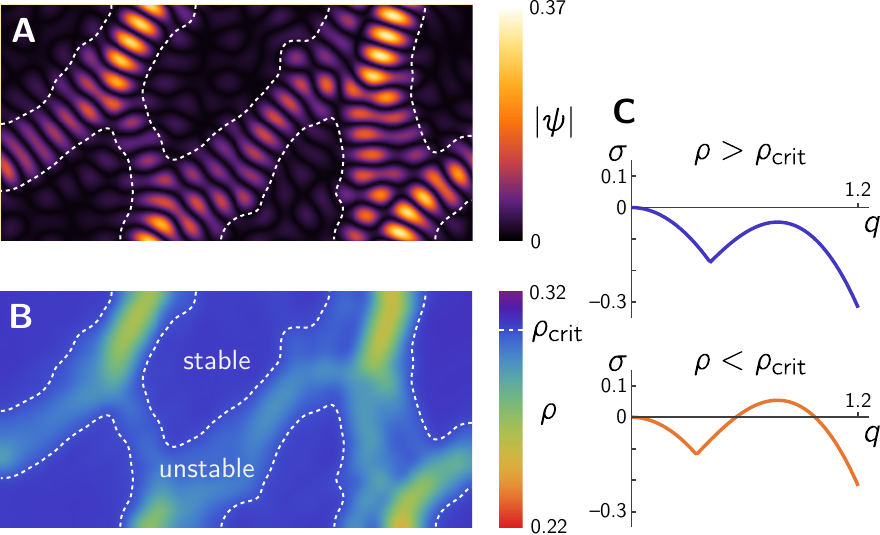}
    \caption{Regional dispersion relations predict localized patterns in the Aranson--Tsimring model \eqref{eq:AT}. (\textit{A}) Snapshot of the order parameter magnitude $\psi$ showing localized patterns. Dashed white line indicates the stability threshold determined from regional dispersion relations. (\textit{B}) Coarse-grained density (Gaussian filter with standard deviation 10). (\textit{C}) Representative dispersion relations in the stable and unstable regimes.
    Domain size: $100\times50$; see Materials and Methods for model details and remaining parmeters.
    }
    \label{fig:AT}
\end{figure}

An example for this third scenario is the Aranson--Tsimring model for pattern formation in vibrated granular media~\cite{Tsimring.Aranson1997} (see Materials and Methods for details). 
In the following, we briefly discuss this model to put our approach into a broader context. 
In particular, this model has been extensively studied using amplitude equations allowing us to connect this mathematical approach to the regional dispersion relations introduced here.
The Aranson--Tsimring model considers a system with a complex order parameter $\psi$ (describing the surface modulation of a vibrated granular layer) which is coupled to a conservation law for the grain density $\rho$ (see \eqref{eq:AT} in Materials and Methods). Near the onset of pattern formation, this coupling gives rise to localized patterns that have been studied using amplitude equations \cite{Winterbottom.etal2005,Winterbottom.etal2008, Matthews.Cox2000}. Figure \ref{fig:AT} and Movie~S9 illustrate how these patterns can be understood in terms of regional dispersion relations.  
For high densities, there are no unstable modes and no patterns form. 
Below a critical density $\rho_c$, a band of unstable modes appears, giving rise to patterns through a supercritical bifurcation.
Indeed, localized patterns appear only where the average regional density is below $\rho_c$ (see Fig.~\ref{fig:AT}B). 
This demonstrates the idea of regional dispersion relations in a nutshell. Moreover, it shows that this approach gives rise to qualitatively similar insights as the technically much more involved amplitude equation formalism.
The conceptual and technical simplicity of regional dispersion relations make this approach readily applicable.
The caveat is that this approach lacks the mathematical rigor of the amplitude equation formalism and requires numerical solutions of the dynamics as a basis.

Because the bifurcation in the Aranson--Tsimring model is supercritical~\cite{Tsimring.Aranson1997}, we can immediately read off from the regional dispersion relation where small-scale patterns will form. 
There is only one pattern type (stripes) and therefore, no additional information is needed to reconstruct the small-scale patterns.
For the Min system considered here, on the other hand, the onset is subcritical (i.e.\ patterns have large amplitude at onset), and the emergence of a band of unstable modes alone does not inform about the pattern type.
To overcome this problem, we used the ``commensurability criterion'' that enabled us to predict the small-scale patterns from regional dispersion relations.
However, this criterion has so far only been shown to hold for the Min-protein system. 
Whether it also applies to other reaction--diffusion systems remains an open question. 
In general, reconstructing subcritical small-scale patterns from large-scale quantities will require that adequate criteria are first identified in simplified settings (such as the ``slice geometry'' used here).
Supercriticality also guarantees that there is no multistability of different pattern types near onset. 
Multistability would lead to hysteresis in the transitions and therefore introduce memory in the system.
As a result, the one-to-one correspondence between the regional dispersion relation and the pattern type that we have used here to reconstruct patterns would be lost.
Handling memory effects in pattern-forming systems remains an open issue that will require the development of new methods, providing an interesting task for future research.

Since conservation laws are ubiquitous in many physical systems, we believe that our approach can be generalized to a broad class of multiscale pattern-forming systems.
For instance, mass conservation is inherent to particle--based active matter systems.
The local particle density controls emergent orientational order, i.e.\ local symmetry breaking \cite{Marchetti.etal2013,Baer.etal2020,Denk.Frey2020}. In turn, orientational order controls mass redistribution due to the particles' self-propulsion.
Thus, the particle density again plays a dual role as a control parameter and a dynamic variable \cite{Denk.Frey2020,Dauchot.Loewen2019,Grossmann.etal2020}.
The dynamic interplay of mass redistribution and orientational order has been shown to give rise to coexistence of different macroscopic order (polar flocks, nematic lanes) and the interconversion between them \cite{Denk.Frey2020}, not unlike the coexistence and interconversion of different patterns we found for the reaction--diffusion system studied in this work.
One way to induce spatial heterogeneities in these systems is to introduce a gradient of signaling chemicals (chemoattractants) that affect the local velocity of active particles.
This would dynamically lead to redistribution of the particle densities on large scales. Since the particle densities, in turn, are themselves control parameters locally, non-trivial multiscale dynamics may emerge in such a setup.
Exploring the effects of such gradients in active matter systems could be therefore an exciting task for future research.

On a broader perspective, our work shows how a linear analysis on small scales, combined with a reduced description for non-linear large scale dynamics (mass redistribution) can be employed to study complex multiscale phenomena.
We believe that our approach can be generalized and applied to other multiscale systems with an underlying conservation law, such as transport processes in porous media, combustion, and cell migration, to name a few examples.

\clearpage

\paragraph{Acknowledgments.} 
E.F. acknowledges support from the Deutsche Forschungsgemeinschaft (DFG, German Research Foundation) through the Collaborative Research Center (SFB) 1032 -- Project-ID 201269156 -- and the Excellence Cluster ORIGINS under Germany’s Excellence Strategy -- EXC-2094 -- 390783311. C.D. acknowledges support from the NWO/OCW Gravitation program NanoFront and the ERC Advanced Grant 883684.

\section*{Materials and Methods}

\subsection*{Mathematical model}
We adopt the Min ``skeleton model'' introduced in Refs.~\cite{Huang.etal2003,Halatek.Frey2012,Halatek.etal2018} which is known to qualitatively reproduce Min patterns \emph{in vivo} and \emph{in vitro} \cite{Halatek.Frey2012,Halatek.etal2018,Brauns.etal2021a}.
The governing equations are given in the main text, Eqs.~[\ref{eq:cyt_dyn}]--[\ref{eq:bc}].
The membrane reactions are
\begin{equation} \label{eq:Min-reactions}
	\mathbf{r} = \left[
	    r_\mathrm{D}^\mathrm{on} - r_\mathrm{E}^\mathrm{on},
	    r_\mathrm{E}^\mathrm{on} - r_\mathrm{DE}^\mathrm{off}
	\right]^{\top},
\end{equation}
with
\begin{subequations}
	\begin{align}
		r_\mathrm{D}^\mathrm{on} &= (k_\mathrm{D} + k_\mathrm{dD} m_\mathrm{d}) c_\mathrm{DT} \,\label{eq:mem-reac-rDon} , \\
		r_\mathrm{E}^\mathrm{on} &= k_\mathrm{dE} m_\mathrm{d} c_\mathrm{E} \, , \\
		r_\mathrm{DE}^\mathrm{off} &= k_\mathrm{de} m_\mathrm{de} \,.
	\end{align}
\end{subequations}
The reaction terms account for MinD attachment and self-recruitment to the membrane, MinE recruitment by MinD, and dissociation of MinDE complexes with subsequent detachment of both proteins into the cytosol, respectively. 
Coupling between cytosol and membrane is established by reactive boundary conditions at the membrane [cf.\ \eqref{eq:bc}]. The boundary fluxes are given by
\begin{equation} \label{eq:Min-fluxes}
    \boldsymbol{f} = \left[
        r_\mathrm{DE}^\mathrm{off},
        -r_\mathrm{D}^\mathrm{on},
        r_\mathrm{DE}^\mathrm{off} - r_\mathrm{E}^\mathrm{on}
    \right]^{\top}\, ,
\end{equation}
which follows from mass conservation.
For analytical caclulations we adapt the following change of variables as it is more convenient: We describe the bulk dynamics of MinD in terms of the variables $c_\mathrm{D}=c_\mathrm{DD}+c_\mathrm{DT}$ and $c_\mathrm{DD}$, i.e. in this case one defines the bulk concentration vector $\mathbf{c} = (c_\mathrm{D},c_\mathrm{DD},c_\mathrm{E})$.
The membrane reaction in~\eqref{eq:mem-reac-rDon} is then slightly modified by substituting $c_\mathrm{DT}=c_\mathrm{D}-c_\mathrm{DD}$, and the boundary fluxes are given by
\begin{equation} \label{eq:Min-fluxes-var-change}
    \boldsymbol{f} = \left[
        -r_\mathrm{D}^\mathrm{on},
        r_\mathrm{DE}^\mathrm{off},
        r_\mathrm{DE}^\mathrm{off} - r_\mathrm{E}^\mathrm{on}
    \right]^{\top} \, .
\end{equation}
The model parameters used in this study are summarized in Table~\ref{tab:parameters}.
\setlength{\tabcolsep}{20pt}
\begin{table}[htb]
\centering
\caption{\textbf{Min model parameters}}
\begin{tabular}{@{} l  c  l @{}}
Parameter & Symbol & Value\\
  \hline	
 Bulk diffusion & $D_c$ & $\SI{60}{\micro m^2 \, s^{-1}}$\\
 Membrane diffusion & $D_m$ & $\SI{0.013}{\micro m^2 \, s^{-1}}$ \\
Average total MinD density & $\bar{n}_\mathrm{D}$ & $\SI{665}{\micro m^{-3}}$ \\ 
Aveage total MinE density & $\bar{n}_\mathrm{E}$ & $\SI{410}{\micro m^{-3}}$ \\ 
Attachment rate & $k_\mathrm{D}$ & $\SI{0.065}{\micro m \, s^{-1}}$\\
MinD recruitment rate  & $k_\mathrm{dD}$ & $\SI{0.098}{\micro m^3 \, s^{-1}}$\\
MinE recruitment rate  & $k_\mathrm{dE}$ & $\SI{0.126}{\micro m^3 \, s^{-1}}$\\
MinDE dissociation rate & $k_\mathrm{de}$ & $\SI{0.34}{s^{-1}}$\\
Nucleotide exchange & $\lambda$ & $\SI{6}{s^{-1}}$\\
\hline
\end{tabular}
\label{tab:parameters}
\end{table}

\subsection*{Numerical simulations}
To investigate the dynamics of the system, we performed 3D finite element (FEM) simulations using the commercially available software \textit{COMSOL Multiphysics v5.6}.
Numerical simulations were performed for a wedge geometry with lateral length $L=\SI{500}{\micro m}$ and bulk height $H(x)$ linearly increasing from $H_0=\SI{5}{\micro m}$ to $H_1=\SI{50}{\micro m}$.
The simulation was initialized with the Min proteins uniformly distributed in the bulk and a small random spatial perturbation around this uniform state.

\subsection*{Homogeneous steady state and dispersion relation}
The homogeneous steady state concentrations, ($\mathbf{c}^*|_{z=0}(H,\bar{n}_\mathrm{D},\bar{n}_\mathrm{E})$, $\mathbf{m}^*(H,\bar{n}_\mathrm{D},\bar{n}_\mathrm{E})$) are obtained from the stationary solutions of Eqs.~[\ref{eq:cyt_dyn}]--[\ref{eq:bc}] together with the mass conservation condition \eqref{eq:total_avg_densities}: 
\begin{equation} \label{eq:local-equilibria-eqs}
    \renewcommand\arraystretch{1.2}
    \left\{
	\begin{array}{rl}
	    \mathbf{r}(\mathbf{c}^*|_{z=0}, \mathbf{m}^*) =& \kern-1.2ex \mathbf{0}, \\
		\boldsymbol{f}\big(\mathbf{c}^*|_{z=0}, \mathbf{m}^*) =& \kern-1.2ex \boldsymbol{\Phi}, \\
		c^*_\mathrm{D}|_{z=0} + (m^*_\mathrm{d} + m^*_\mathrm{de}) / H =& \kern-1.2ex \bar{n}_\mathrm{D}, \\
		c^*_\mathrm{E}|_{z=0} + m^*_\mathrm{de}/H =& \kern-1.2ex \bar{n}_\mathrm{E},
	\end{array}
	\right.
\end{equation}
where $\boldsymbol{\Phi}$ denotes the steady state fluxes at the membrane, given by:
\begin{subequations}
\begin{align}
    \mathbf{\Phi} &= \left[0,\phi,0\right]^{\top}, \\
    \phi &:= \sqrt{D_c \lambda} \, \tanh\left(\sqrt{\lambda/D_c}\,H\right)\, c^*_\mathrm{DD}|_{z=0}.
\end{align}
\end{subequations}
A concise derivation of these equations and how they can be solved is provided in the Supplementary Information. 
For a thorough presentation of the linear stability analysis of the Min system in a 2D rectangular geometry we refer to the Supplementary Informations of Refs.~\cite{Halatek.Frey2018} and~\cite{Brauns.etal2021a}.

\subsection*{Operators for spatial averaging}
The operators used throughout this study to calculate mean values of densities on the membrane and in the cytosol are defined as follows:
\begin{subequations}
\begin{align}
    \langle m \rangle_\mathcal{S} &:= |\mathcal{S}|^{-1} \int_\mathcal{S} \mathrm{d} x \mathrm{d} y \, m, \\
    \langle c \rangle_\mathcal{V} &:= |\mathcal{V}|^{-1} \int_\mathcal{S} \mathrm{d} x \mathrm{d} y \int_0^{H(x)} \mathrm{d} z \, c, \\ 
    \langle \cdot \rangle_y &:= \frac{1}{L} \int_0^L \mathrm{d} y \, (\cdot) , \\
    \langle \cdot \rangle_{y,z} &:= \frac{1}{\, H(x)} \int_0^{H(x)} \mathrm{d} z \, \langle \cdot \rangle_y, 
\end{align}
\end{subequations}
where the membrane surface area and the bulk volume for the wedge geometry are explicitly given by $|\mathcal{S}| = L^2$ and $|\mathcal{V}| = L^2 \, (H_0 + H_1)/2$.

\subsection*{Instantaneous total densities \emph{at} the membrane}
Since only cytosolic proteins in close proximity to the membrane participate in the nonlinear dynamics at the membrane, we define instantaneous total densities at the membrane:
\begin{subequations}
\begin{align}
    \tilde{n}_\mathrm{D}(x,y,t) &:= \frac{1}{H(x)} (m_\mathrm{d} + m_\mathrm{de}) + c_\mathrm{D}|^{}_{z=0} \,, \\
    \tilde{n}_\mathrm{E}(x,y,t) &:= \frac{1}{H(x)} m_\mathrm{de} + c_\mathrm{E}|^{}_{z=0} \,.
\end{align}
\end{subequations}
We further averaged these densities along the $y$--direction to obtain the the slice-averaged total densities $\langle \tilde{n}_\mathrm{D,E} \rangle_y (x,t)$.
Note that the length of a slice is much larger than the typical pattern wavelength, which also permits to approximate the slice-averaged mass at the membrane by the vertically averaged mass:
$\langle \tilde{n}_i \rangle_{y}(x,t) \approx \langle n_i \rangle_{y,z}(x,t)$ (see Ref.~\cite{Halatek.Frey2018}).
This is because the local deviations $\tilde{n}_i - \langle n_i \rangle_z$ largely cancel when averaging over the pattern wavelength.

\subsection*{Mass redistribution dynamics}
Here, we provide more details on the derivation of the mass redistribution dynamics~\eqref{eq:mass-redistri-exact}. For specificity, we present the calculation for MinD. The calculation for MinE works along the same lines.
Our starting point is the slice averaged total MinD density:
\begin{equation}
    \langle n_\mathrm{D} \rangle_{y,z}(x,t) := \frac{1}{\, H(x)} \left\langle
        m_\mathrm{d} + m_\mathrm{de} +
        \int_0^{H(x)} \mathrm{d} z \, c_\mathrm{D}  
    \right\rangle_{y}.
\end{equation}
The time evolution of this quantity then follows from~\eqref{eq:cyt_dyn} and~\eqref{eq:mem_dyn}:
\begin{multline} 
\label{eq:slice-mass-dyn}
    H(x) \, \partial_t \langle n_\mathrm{D} \rangle_{y,z}(x,t) =
    D_m \partial_x^2 \langle m_\mathrm{d} + m_\mathrm{de} \rangle_y\\
    + D_c \partial_z \langle c_\mathrm{D} \rangle_y \big|_{z = H(x)}
    +
    \int_0^{H(x)} \mathrm{d} z \, D_c \partial_x^2 \langle c_\mathrm{D} \rangle_y,
\end{multline}
where we used the reactive boundary condition~\eqref{eq:bc} to rewrite the integral:
\begin{align}
    \int_0^{H(x)} \mathrm{d} z \, D_c \partial_z^2 c_\mathrm{D} &= 
    D_c \partial_z c_\mathrm{D} \big|_{z = H(x)} - 
    D_c \partial_z c_\mathrm{D} \big|_{z = 0}\nonumber \\
    &= D_c \partial_z c_\mathrm{D} \big|_{z = H(x)} + r_\mathrm{DE}^\mathrm{off}-r_\mathrm{D}^\mathrm{on}.
\end{align}
Note that due to mass-conservation the reaction terms at the membrane cancel.

Since the system is closed, the boundary condition at the inclined top surface of the wedge reads 
$\mathbf{n} \cdot \nabla c_\mathrm{D}|_{z = H(x)} = 0$, where $\mathbf{n} \propto (-\partial_x H,0,1)$ is the outward normal vector at the top surface.
Writing out the boundary condition explicitly, we find that:
\begin{equation}
    \partial_z  c_\mathrm{D} |_{z = H(x)} = (\partial_x H) \, \partial_x  c_\mathrm{D}|_{z = H(x)}.
\end{equation}
To proceed, we substitute the relation above into~\eqref{eq:slice-mass-dyn} and slightly rewrite the resulting equation by applying the chain rule:
\begin{equation}
\label{eq:mass-redist-exact-methods}
    H(x) \, \partial_t \langle n^{}_\mathrm{D} \rangle_{y,z}(x,t) = D_m \partial_x^2 \langle m_\mathrm{d} + m_\mathrm{de} \rangle_y
    +
    \partial_x \underbrace{\int_0^{H(x)} \mathrm{d} z \, 
        D_c \partial_x \langle c^{}_\mathrm{D} \rangle_y
    }_{\displaystyle
      =: - H(x)J_\mathrm{D}(x)
    }.
\end{equation}
Here, the first term describes diffusion of the averaged membrane concentrations.
The integral on the right describes diffusion of the averaged cytosolic densities, where we defined the diffusive flux $J_\mathrm{D}=-D_c\langle \partial_x c_D \rangle_{y,z}$.
The factor $H(x)$ in the cytosolic diffusion term accounts for the increasing area of the slice along the positive $x$--direction.

Since protein diffusion on the membrane is much smaller than cytosolic diffusion $D_m \ll D_c$ \cite{Loose.etal2011,Meacci.etal2006}, one can neglect membrane diffusion to arrive at the result shown in the main text (\eqref{eq:mass-redistri-exact}).
For completeness, note that~\eqref{eq:mass-redist-exact-methods} (without membrane diffusion) can be recast as
\begin{align}
    \partial_t \langle n^{}_\mathrm{D} \rangle_{y,z}(x,t)
    &\approx
    \frac{1}{H(x)} \, \partial_x \int_0^{H(x)} \mathrm{d} z \, D_c \partial_x \langle c^{}_\mathrm{D} \rangle_y, \notag \\
    &=
    D_c \partial_x \langle \partial_x c^{}_\mathrm{D} \rangle_{y,z} + D_c \frac{\partial_x H(x)}{H(x)} \langle \partial_x c^{}_\mathrm{D} \rangle_{y,z},
\end{align}
which is the form given in~\eqref{eq:mass-redistri-exact-driftterm} in the main text.

\subsection*{Machine-learning based pattern classification}
We used the pixel classifier provided by the software \emph{ilastik} \cite{Berg.etal2019}. The classifier was trained based on a few representative snapshots, by manually marking areas where the pattern type (no pattern, chaos, standing wave, or traveling wave) is easily identified by visual inspection. The trained classifier then yields probabilities for each pattern type at each pixel. The classifier was applied to snapshots in 20~s intervals. This data was then downsampled and averaged over slices to yield an $x$--$t$ space time map of pattern probabilities. To render the kymograph in Fig.~\ref{fig:pattern-type-kymograph}D each pixel was colored based on the most probable pattern.

\subsection*{Aranson--Tsimring model}

As a second example we briefly discuss a phenomenological model for pattern formation in vibrated granular media introduced in~\cite{Tsimring.Aranson1997}. This model, which we call Aranson--Tsimring model in the following, couples a Ginzburg--Landau-type equation~\cite{Aranson.Kramer2002} for the complex order parameter $\psi$ to a conservation law for the density $\rho$: 
\begin{subequations} \label{eq:AT}
\begin{align}
    \partial_t \psi &= \gamma \bar{\psi} - (1-i \omega) \psi + (1 + i b) \nabla^2 \psi - |\psi|^2 \psi - \rho \psi, \label{eq:AT-psi}\\
    \partial_t \rho &= \beta \nabla^2 \rho + \alpha \nabla \cdot (\rho \nabla |\psi|^2), \label{eq:AT-rho}
\end{align}
\end{subequations}
where $\bar{\psi}$ denotes the complex conjugate of $\psi$.
The coupling is such that increasing the density $\rho$ suppresses the instability in 
\eqref{eq:AT-psi} while gradients in the amplitude $|\psi|$ drive mass redistribution away from high amplitude regions (second term in \eqref{eq:AT-rho}). This feedback loop amplifies heterogeneities in the density and gives rise to localized patterns.
These patterns have been studied in detail using amplitude equations in \cite{Winterbottom.etal2005,Winterbottom.etal2008}. Moreover, in Ref. \cite{Matthews.Cox2000} it was shown that the system \eqref{eq:AT} appears as the amplitude equation for a mass-conserving version of the classical Swift--Hohenberg--Turing equation \cite{Cross.Hohenberg1993,Dawes.2016}. The reason for this is that the conserved density appears as a second hydrodynamic variable in addition to the pattern amplitude.

A linear stability analysis shows that the system \eqref{eq:AT} has a short wavelength instability when $b \omega - 1 - \rho_0 > 0$ and $\gamma > \gamma_c = (\omega + b(1+\rho_0))/\sqrt{1+b^2}$, where $\rho_0$ denotes the average density. 
Following Ref.~\cite{Winterbottom.etal2008}, we set parameters $b = 1, \omega = 2.5, \alpha = 1.3, \beta = 0.3, \rho_0 = 0.3$. Localized patterns are found near the instability threshold, so we set $\gamma = 1.001 \gamma_c$ for the simulation shown in Fig.~\ref{fig:AT} and Movie~S9.

\subsection*{Preparation of the wedge flow cell}
The microfluidic wedge chambers were prepared using two rectangular cover slips (bottom one of dimensions $22$/$\SI{50}{\milli m}$, and top one of dimensions $5$/$\SI{30}{\milli m}$). 
Close to one of the short edges of a top glass a tiny inlet hole was drilled using a sandblaster. 
Cover slips were cleaned in $1$ M KOH for $1$ h followed by a methanol bath for $10$ min in a sonicator bath. 
Surfaces of the cover slips were activated with oxygen plasma for $20$ s, using oxygen plasma PREEN I (Plasmatic System, Inc.) with a O2 flow rate of $1$ SCFH. 
Furthermore, a small PDMS slab with a $\SI{0.3}{\milli m}$ hole was attached on to the top glass slide, such that it matches the hole in the PDMS glass slide and a metal connector was inserted in the hole for connecting the syringe pump. 
Tilt of the top glass slide was achieved by placing a piece of aluminum foil between the top and bottom slide at the end, with the largest height between top and bottom at the side of the inlet. 
At the opposite side with the smallest distance between top and bottom slide, $\SI{2}{\micro m}$ polystyrene beads that were deposited on the bottom slide provided an outlet and prevent a collapse of the top and bottom slides (see Fig.~\ref{fig:exp-setup}). 
The lateral sides of the microchamber were sealed with a two-component epoxy resin leaving the short edge at the low height-side open for liquid flow (Fig.~S4). The microfluidic cell was then filled with a solution of small unilamellar vesicles (SUVs) through an injection tube at the inlet of the PDMS slab and incubated for 30 min at \SI{30}{\celsius}--yielding full lipid membrane coverage of the bottom and top slides. SUVs were prepared as described in Ref.~\cite{Brauns.etal2021a}. Subsequently, the flow cell was thoroughly washed with a buffer to remove excess SUVs and Min protein experiments were started. 

\subsection*{Observation of Min patterns}
We purified the Min proteins based on the method proposed in Ref.~\cite{Caspi.Dekker2016}. 
Injection of Min proteins into the flow cells was performed through a syringe pump containing a solution of $\SI{0.8}{M}$ MinD, $\SI{0.2}{\milli M}$ MinD-Cy3, $\SI{0.8}{\milli M}$ MinE, $\SI{0.2}{\milli M}$ MinE-Cy5, $\SI{5}{\milli M}$ ATP, $\SI{4}{\milli M}$ phosphoenolpyruvate, $\SI{0.01}{\milli g/\milli l}$ pyruvate kinase, $\SI{25}{\milli M}$ Tris-HCl (pH 7.5), $\SI{150}{\milli M}$ KCl and $\SI{5}{\milli M}$ MgCl2. 
To ensure that all of the buffer solution in the microdevice is replaced by the protein solution, we chose a volume of the protein solution that was 50 times larger than the volume in the microdevice. 
During the filling process of the microdevice, the enire solution was rapidly injected (in  5 s) to prevent protein accumulation on the membrane.

For the generation of the fluorescence images, we used the following equipment: Olympus IX-81 inverted microscope equipped with an Andor Revolution XD spinning disk system with FRAPPA, illumination and detection system Andor Revolution and Yokogawa CSU X1, EM-CCD Andor iXon X3 DU897 camera, motorized x-y stage and a z-piezo stage, using a 20x objective (UPlansApo, NA 0.85, oil immersion). Imaging of MinD-Cy3 and MinE-Cy5 was performed with laser spectral lines at 561 nm and 640 nm, respectively, and we further used a 617/73 band-pass filter as well as a 690 long-pass filter. We imaged several uniformly sized regions at intervals of 30 s or 60 s along the lateral length of the wedge setup. To exclude membrane imperfections that may have arisen during preparation, we also imaged the membrane using the spectral line at $\SI{491}{\nano m}$ and a 525/50 band-pass filter.

\subsection*{Image sequence processing}
We processed the fluorescence images using the following software packages: Andor iQ3 v3.1, ImageJ 1.52j, and custom written Matlab 2016a scripts. For better visualization, we additionally applied background correction and filtering of artifacts. In detail, these were carried out as follows: For the generation of the movies, each frame was first corrected for fluorescence bleaching (max. 20 \% decay of the intensity for long movies) by normalizing to the mean intensity of the respective frame. Then, we generated two different modifications of the images: First, we averaged out all transient features (i.e., patterns) in the frames to obtain ‘static background’-images which we shall call \texttt{Imstat}. Second, we smoothed out the images, determined the average of all movie frames, and normalized the corresponding result with respect to its maximum. This way, we obtained an ‘illumination correction’ image \texttt{Imillum}. In the final step, each frame \texttt{Immovie} was corrected according to the rule \texttt{Imcorrected} = (\texttt{Immovie} - \texttt{Imstat})/\texttt{Imillum}. On one hand, this ensures that irregularities in each image are suppressed, and on the other hand, the intensity amplitudes at the edges becomes comparable with the values at the center of the image.

\newpage


\end{document}